\documentclass[traditabstract]{aa}
\usepackage[utf8]{inputenc}
\usepackage{amsmath}
\usepackage{rotating}
\usepackage{natbib}
\usepackage{graphicx}
\usepackage{epsfig}
\usepackage{txfonts}
\usepackage{color}
\usepackage{multirow}
\usepackage{amssymb}
\usepackage{epsfig}
\bibpunct{(}{)}{;}{a}{}{,} 
\usepackage[plainpages=false]{hyperref}
\title{\textsl{RXTE} observations of the 1A~1118$-$61 in an outburst, and the discovery of a cyclotron line.}
\author{V. Doroshenko\inst{1},  S. Suchy\inst{2}, A. Santangelo\inst{1}, R. Staubert\inst{1}, I. Kreykenbohm\inst{3,4}, R. Rothschild\inst{2},	K.~Pottschmidt\inst{5,6}, J.~Wilms\inst{3,4}}
\institute{
Institut für Astronomie und Astrophysik, Sand 1, 72076 Tübingen, Germany \and
University of California, San Diego CASS, M/C 0424 9500 Gilman Drive La Jolla, CA 92093-0424, USA\and
Dr. Karl Remeis-Sternwarte, Sternwartstrasse 7, 96049 Bamberg, Germany \and
Erlangen Centre for Astroparticle Physics (ECAP), Erwin-Rommel-Strasse 1, 91058 Erlangen, Germany \and
CRESST and NASA Goddard Space Flight Center, Astrophysics Science Division, Code 661, Greenbelt, MD 20771, USA \and
CSST, University of Maryland, Baltimore County, 1000 Hilltop Circle, Baltimore, MD 21250, USA
}

\begin{document}

\bibliographystyle{aa}

\abstract{
We present the analysis of \textsl{RXTE} monitoring data obtained during
the January~2009 outburst of the hard X-ray transient 1A~1118$-$61. Using
these observations the broadband (3.5--120 keV) spectrum of the source was
measured for the first time ever. We have found that the broadband
continuum spectrum of the source is similar to other accreting pulsars and
is well described by several conventionally used phenomenological models.
We have discovered that regardless of the
applied continuum model, a prominent broad absorption feature at $\sim$55\,keV
is observed. We interpret this feature as a cyclotron resonance scattering
feature (CRSF). The observed CRSF energy is one of the highest known and
corresponds to a magnetic field of B$\sim4.8\times10^{12}$\,G in the scattering
region. Our data also indicate the presence of an iron emission line presence that has not been previously reported for 1A~1118$-$61.
Timing properties of the source, including a
strong spin-up, were found to be similar to those observed by
\textsl{CGRO}/BATSE during the previous outburst, but the broadband
capabilities of \textsl{RXTE} reveal a more complicated energy dependency
of the pulse-profile.} 

\authorrunning{V. Doroshenko et al.}

\maketitle

\section{Introduction} 
\label{sec:introduction}
The hard X-ray transient 1A~1118$-$61 was first discovered during an
outburst in 1974 by the \textsl{Ariel-5} satellite \citep{Eyles:1975p3466}.
The outburst lasted for $\sim$10 days and no flux could be observed
afterwards. Pulsations with a period of 405.6\,s were observed by
\citet{Ives:1975p3476} and were initially interpreted as the orbital period
of two compact objects. It was suggested by \cite{fabian75} that the
observed period was due to a slow rotation of the neutron star. The optical
counterpart was identified as the Be-star He 3$-$640/Wray 793 by
\citet{Chevalier:1975p3481} and classified as an O9.5IV-Ve star with strong
Balmer emission lines and an extended envelope by
\citet{JanotPacheco:1981p3486}. The distance was estimated to be
$5\pm2$\,kpc \citep{JanotPacheco:1981p3486}. The classification and
distance was confirmed by \citet{Coe:1985p3489} by UV observations of
the source. The X-ray spectrum of the pulsar was fitted with a power law
with a photon index of $\Gamma \sim 1 $, with a possible spectral softening
to $\Gamma \sim 0.9$ during the
peak of the outburst (significant at $1\sigma$ confidence level). 

A second outburst occurred in 1992 and was observed by
\textsl{CGRO}/BATSE \citep{Coe:1994p3488}. The measured peak pulsed flux was $\sim
150$ mCrab for the 20--100\,keV energy range, similar to the 1974 outburst
\citep{Coe:1994p3488,maraschi}. It was followed by a period
of elevated emission $\sim 25$ days after the main outburst. This lasted for $\sim
30$~days \citep[see][Fig.1]{Coe:1994p3488}. Pulsations with a period of 406.5\,s
were detected up to 100\,keV and the pulse profile showed a single, broad
peak, asymmetric at lower (20--40\,keV) energies. A spin-up of 0.016\,s/day
was observed during the decay of the outburst. The pulsed spectrum was
described with a single-temperature optically-thin, thermal bremsstrahlung
model with a temperature of ($15.1 \pm 0.5$)\,keV for the main outburst and
($18.5 \pm0.9$)\,keV during the elevated emission. Multi-wavelength observations
revealed a strong correlation between the $H_\alpha$ equivalent width and
the X-ray flux, which allowed \cite{Coe:1994p3488} to conclude that
expansion of the circumstellar disk of the optical companion was mainly
responsible for the increased X-ray activity. The periastron passage would then
trigger an outburst if enough matter had accumulated in the system. This
conclusion was supported by the pulsations with a period of
$\sim409$\,s, which were also detected in quiescence \citep{quescence}.

The source remained in quiescence until 2009~January~4, when a third outburst was
detected by \textsl{Swift} \citep{Mangano09_GCN8777}. Pointed observations
with \textsl{Swift}/XRT allowed the detection of pulsations with a period
of $407.68\pm0.02$\,s reported later by \cite{swift_atel}. The complete
outburst was regularly monitored with \textsl{RXTE}. INTEGRAL observed the
source after the main outburst and observed flaring activity $\sim 30$
days after the main burst \citep{intatel}. \textsl{Suzaku} observed
1A~1118$-$61 twice, once during the peak of the outburst and also $\sim
20$\,days later when the flux returned to its previous level (S. Suchy, in
preparation). 

We report here on the monitoring observations by \textsl{RXTE}. A
timing analysis to determine the pulse period of the source and the rate of
the observed strong spin-up was carried out. Pulse phase averaged spectral
analysis revealed an absorption feature at $\sim55$\,keV,
confirmed independently with \textsl{Suzaku} (S. Suchy, in preparation). We
interpret this feature as a cyclotron resonance scattering feature (CRSF), which is
observed for the first time in this source.

\section{Observations}
\begin{figure}[t]
	\centering
		\includegraphics{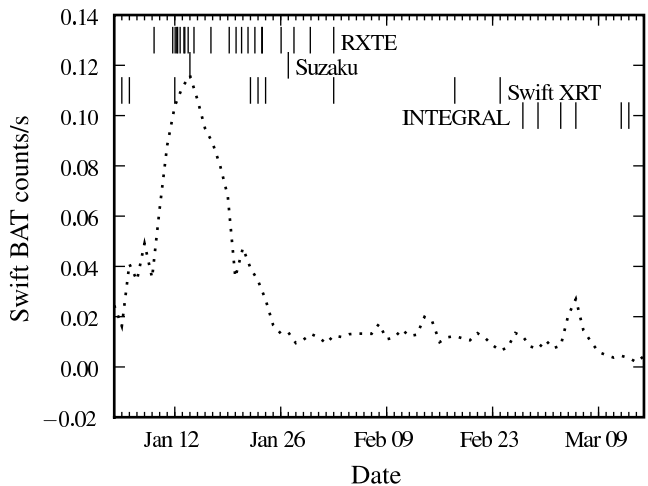}
    \caption{\textsl{Swift/BAT} daily light-curve of the outburst in 2009~January (dotted line). Observation times by various missions
    are indicated with vertical lines.} 
\label{fig:bat}
\end{figure}

The outburst was observed by multiple X-ray missions. The source
light-curve as observed by \textsl{Swift/BAT} with marked observation times
by various satellites is presented in Fig.\ref{fig:bat}. The source flux
peaked on 2009~January~14, (\textsl{Swift/BAT} countrate 0.12 counts/s
corresponding to $\sim500$mCrab in the 15--50\,keV energy range) and slowly
decreased afterwards. The source was regularly monitored with \textsl{RXTE} between January 10
and 2009~February~4, with a total exposure of 86\,ks (PCA) and
dead-time-corrected live time of 29\,ks (HEXTE-B). The
data from the HEXTE-A detector were not used as it was not rocking during the
entire observation and the background could not be properly estimated. The
data were reduced with the HEASOFT version 6.8 and a set of calibration files version 20091202. The spectral modeling was performed
with the \textsl{XSPEC} package version 12.5.1n. 
The energy range 3.5--25\,keV was
used for the PCA and 18--120\,keV for the HEXTE spectra. Based on an analysis of
recent Crab observations performed during the same time frame (2008~December~16, -- 2009~March~11, proposal ID 
P94802), a systematic of 0.5\% was determined for the PCA data. No systematic error was
required for the HEXTE data. 

\section{Timing analysis} 
\begin{figure}[t]
	\centering
		\includegraphics{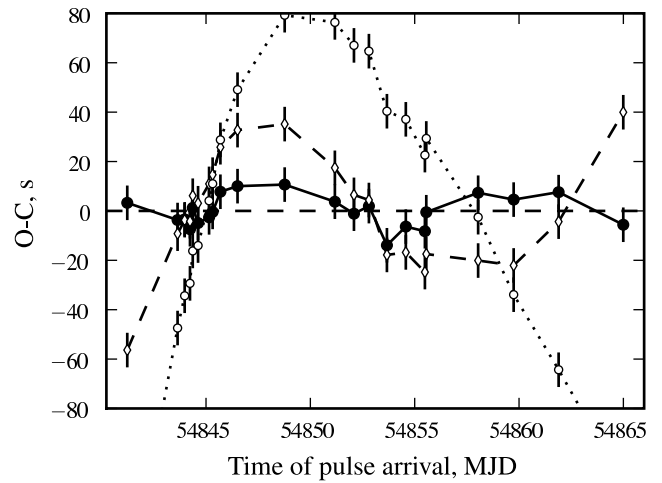}
    \caption{Best-fit residuals for the pulse arrival times determined using
    \textsl{RXTE} PCA lightcurve (i.e. difference between pulse arrival times
    as observed (O) and as calculated (C) with an assumed pulse period and derivative).
    Solid, dashed and dotted lines correspond to fits including the
    pulse period derivatives up to second, first and none (constant period)
    respectively.}
	\label{fig:toa}
\end{figure}

To determine the pulse period of 1A~1118$-$61 the PCA light-curve covering
the complete observation in the 3--50\,keV energy range was used. We
searched for the pulse period and pulse period derivative using the phase
connection technique \citep{staubert2009}. Our best-fit results give
$P_\mathrm{spin}$=407.719(9)\,s,
$\dot{P}_\mathrm{spin}=-4.6(2)\times10^{-7}\,\rm{s}\,\rm{s}^{-1}$,
at a folding epoch of MJD\,54841.62 (the uncertainties given in parenthesis are at $1\sigma$ confidence level and
refer to the last digit given). The best-fit residuals for pulse arrival
times are presented in Fig.~\ref{fig:toa}. Note that the obtained parameters
characterize the pulse period evolution during the \textsl{RXTE}
observations only and are not consistent with the pulse period value provided by
\cite{swift_atel}. This inconsistence is anticipated, because the luminosity
and hence the accretion rate were significantly higher during the
\textsl{RXTE} observations. The spin-up rate of the neutron star depends on
the accretion rate, and therefore extrapolating the timing solution obtained
close to the peak of the outburst (with a stronger spin-up) to the
beginning of the outburst (where the spin-up rate is expected to be
significantly lower) gives a longer pulse period than can be directly measured at the time.
Note also that our solution does not account for the Doppler delays caused
by orbital motion, because the parameters of the orbit are unknown and
our data do not allow us to find an unambiguous solution for the orbit.
The estimated value for an orbital period from ``pulse period''--``orbital period''
diagram \cite{corbet} lies in the 400--800\,d range, which is much longer than the span
of our data. The intrinsic spin-period evolution is expected to be complicated,
so it is difficult to separate it from the effects of the orbital motion. 
Moreover, \cite{Coe:1994p3488} suggested that the orbit may be 
almost circular, and the outbursts may occur at any orbital phase, so
a possible orbital period value derived from the comparison of outburst times is 
also potentially ambiguous.

Both the spin-up rate $\sim0.04$\,s\,d$^{-1}$ and flux are somewhat higher
(by a factor of $\sim$2--3) than reported by \citet{Coe:1994p3488} for the
previous outburst observed with \textsl{CGRO}/BATSE. The stronger spin-up measured for the current outburst
is likely caused by the higher flux and consequently higher accretion rate (although it is
difficult to directly compare BAT and BATSE fluxes, because
BATSE measured only pulsed flux and the energy ranges are slightly
different). A potential difference in the orbital phase during the two observations
may also be responsible for the difference in the observed spin-up rate. 

\begin{figure*}
	\centering
		\includegraphics{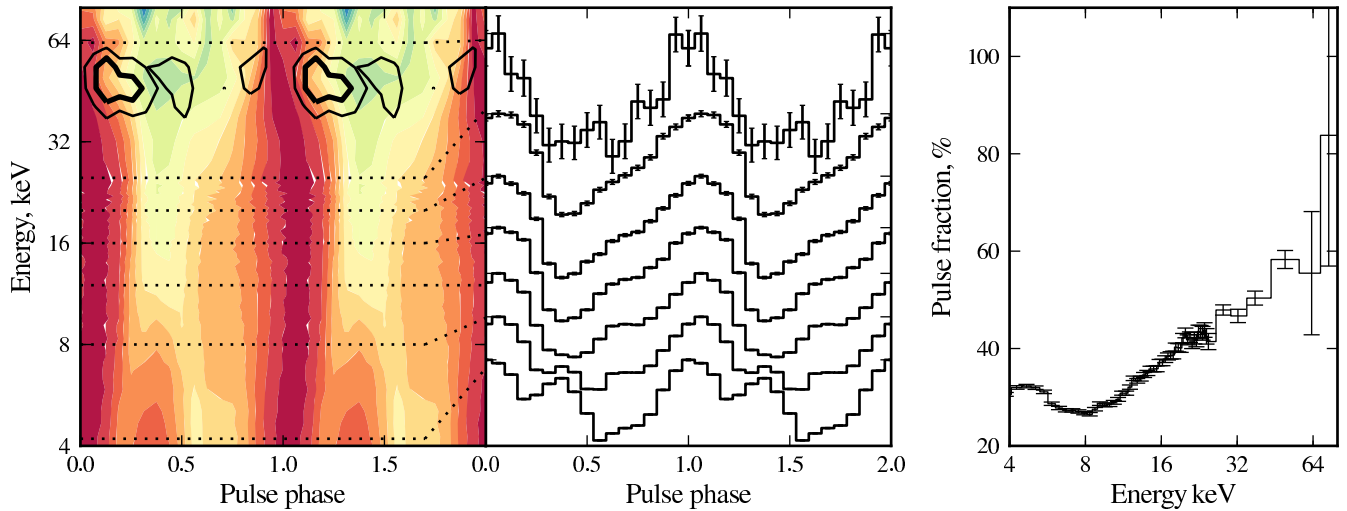}
    \caption{ 
    Normalized ``pulse phase''-``energy'' matrix using PCA (below 22\,keV) and HEXTE (above
	    22\,keV) data and pulse profile evolution with energy. 
    A slice at a constant energy gives a background-subtracted
    pulse-profile normalized to unity at the pulse maximum (shown in the middle pane, the pulse
profiles are shifted with respect to each other to avoid confusion). 
    Contours represent $2\sigma$ and $3\sigma$ significance levels for the
    absorption feature in the residuals to the fit with \texttt{CompTT} model
    modified by photoelectric absorption and emission line at 6.4\,keV.
	The pulse fraction, defined as $(A_{max}-A_{min})/(A_{max}+A_{min})$ is shown as a function of energy in the right panel.}
    \label{fig:enphase}
\end{figure*}

A set of pulse profiles in several energy ranges was constructed with
the determined period. The pulse profile significantly changes with 
energy as shown in Fig.~\ref{fig:enphase}. At energies below 10\,keV the pulse
profile has two peaks. The secondary peak amplitude decreases towards higher
energies, disappearing above 10\,keV. A shoulder appears on the
other side of the main peak at about the same energy. The pulse profile becomes single peaked and gradually more symmetric and
narrow at higher energies. A similar behavior was observed previously with
\textsl{CGRO}/BATSE \citep{Coe:1994p3488}. The pulse fraction, defined as
$(A_{max}-A_{min})/(A_{max}+A_{min})$ increases with energy in a similar way to
other accreting pulsars, except for a drop at around 8\,keV, where the
second peak disappears (see Fig.\ref{fig:enphase}). 

\section{Spectral analysis}
\begin{figure*}[t]
		\includegraphics{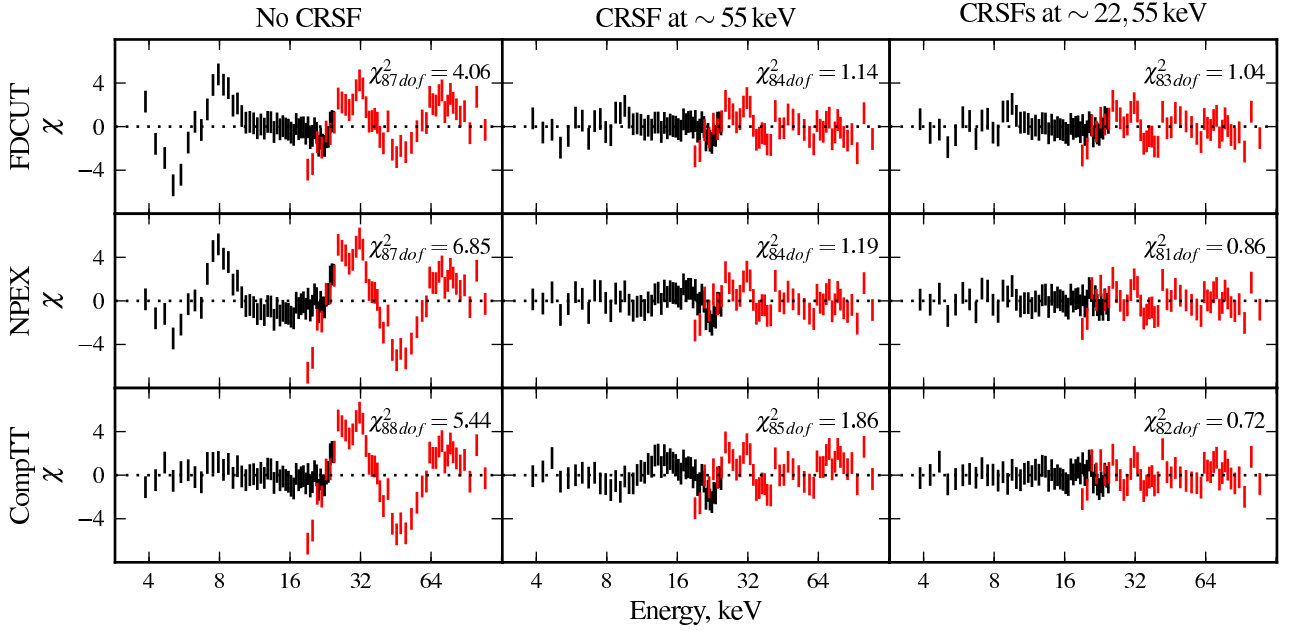}
 \caption{Best-fit residuals for various continuum models of 1A~1118$-$61. A line-like
emission feature at $\sim8$\,keV was modeled with a narrow line with Gaussian profile as described in text. The
 absorption feature at $\sim$23\,keV was included for
 PCA only, to account for the similar residuals in the Crab spectrum
 (see text).}
	\label{fig:spec}
\end{figure*}
\begin{table*}[t]
	\begin{center}
	\begin{tabular}{lllllll}
                                 Parameter &                    \texttt{FDCUT} &                           &                     \texttt{NPEX} &                            &         \texttt{CompTT}$_\mathrm{fix23}$ &        \texttt{CompTT}\\
\hline
                   $N_\mathrm{H}$\,\tablefootmark{(e)} &               $4.9_{-0.6}^{+0.6}$ &                           &               $3.5_{-0.4}^{+0.4}$ &                            &               $0.0_{-0.0}^{+0.4}$ &    $0.2_{-0.2}^{+0.5}$\\
             $E_{22}$\,\tablefootmark{(b)} & \textsl{22.76}\tablefootmark{(a)} &                           & \textsl{22.76}\tablefootmark{(a)} &                            & \textsl{22.76}\tablefootmark{(a)} &   $23.5_{-0.8}^{+1.5}$\\
        $\sigma_{22}$\,\tablefootmark{(b)} &   \textsl{1.8}\tablefootmark{(a)} &                           &   \textsl{1.8}\tablefootmark{(a)} &                            &   \textsl{1.8}\tablefootmark{(a)} &    $4.6_{-1.1}^{+1.9}$\\
                               $\tau_{22}$ &            $0.02_{-0.01}^{+0.01}$ &                           &           $0.032_{-0.01}^{+0.01}$ &                            &            $0.05_{-0.01}^{+0.01}$ & $0.09_{-0.02}^{+0.04}$\\
            $E_{\mathrm{cyc}}$\,\tablefootmark{(b)} &              $55.1_{-1.5}^{+1.6}$ &                           &              $55.2_{-1.5}^{+1.6}$ &                            &              $52.9_{-1.4}^{+1.7}$ &   $55.5_{-2.1}^{+2.5}$\\
       $\sigma_{\mathrm{cyc}}$\,\tablefootmark{(b)} &              $10.4_{-1.0}^{+1.1}$ &                           &              $11.8_{-1.1}^{+1.2}$ &                            &              $10.2_{-1.2}^{+1.5}$ &   $13.3_{-2.0}^{+2.2}$\\
                              $\tau_{cyc}$ &               $0.8_{-0.1}^{+0.1}$ &                           &               $0.9_{-0.1}^{+0.2}$ &                            &               $0.6_{-0.1}^{+0.1}$ &    $0.9_{-0.2}^{+0.3}$\\
                                  $\Gamma$ &            $0.73_{-0.06}^{+0.05}$ &                $\Gamma_1$ &            $0.16_{-0.03}^{+0.03}$ &                            &                                   &                       \\
            $E_{\mathrm{cut}}$\,\tablefootmark{(b)} &              $16.5_{-2.9}^{+2.5}$ &                $\Gamma_2$ &                            $-2.0$ &                     $\tau$ &               $6.0_{-0.1}^{+0.1}$ &    $6.0_{-0.2}^{+0.2}$\\
           $E_{\mathrm{fold}}$\,\tablefootmark{(b)} &              $12.0_{-0.5}^{+0.5}$ &  $A_2$\tablefootmark{(c)} &            $0.16_{-0.02}^{+0.02}$ & $T_0$\,\tablefootmark{(b)} &            $1.47_{-0.03}^{+0.02}$ & $1.44_{-0.05}^{+0.04}$\\
                                           &                                   & $kT$\,\tablefootmark{(b)} &               $7.9_{-0.3}^{+0.3}$ &                            &               $7.2_{-0.2}^{+0.2}$ &    $7.7_{-0.3}^{+0.4}$\\
             $E_{\mathrm{Fe}}$\,\tablefootmark{(b)} &               $6.5_{-0.1}^{+0.1}$ &                           &            $6.45_{-0.08}^{+0.07}$ &                            &             $6.4_{-0.09}^{+0.02}$ &  $6.4_{-0.09}^{+0.02}$\\
        $\sigma_{\mathrm{Fe}}$\,\tablefootmark{(b)} &                          $\le0.4$ &                           &                          $\le0.3$ &                            &                          $\le0.3$ &               $\le0.3$\\
               $A_{\mathrm{Fe}}$\tablefootmark{(d)} &                     $3_{-1}^{+3}$ &                           &               $3.2_{-0.7}^{+0.7}$ &                            &               $2.7_{-0.3}^{+0.3}$ &    $2.7_{-0.3}^{+0.3}$\\
               $A_{\mathrm{Cu}}$\tablefootmark{(d)} &                     $2_{-2}^{+2}$ &                           &                 $1_{-0.3}^{+0.3}$ &                            &                                   &                       \\
$A_{\Gamma,1,\texttt{CompTT}}$\,\tablefootmark{(c)} &                    $22_{-1}^{+2}$ &                           &              $10.7_{-0.7}^{+0.7}$ &                            &              $10.1_{-0.2}^{+0.2}$ &    $9.8_{-0.3}^{+0.3}$\\
                          $\chi_{\mathrm{red}/\mathrm{dof}}$ &                           1.04/83 &                           &                           0.85/83 &                            &                           1.08/84 &                0.72/82\\

	\end{tabular}                                                                                                                       
	\end{center}                                                                                                                        
 \caption{Best-fit results for different models. An emission line (\textsl{Cu})
 with energy and width fixed at 8.04\,keV and 0.01\,keV was added to
 \texttt{FDCUT} and \texttt{NPEX} models. All models include also an absorption
 like feature at $\sim23$\,keV for PCA to account for similar
 residuals seen in Crab spectra (with the line energy and width fixed to those
 obtained from Crab fits). For \texttt{CompTT} model the $\chi^2$ may be
 improved by allowing the line parameters to vary (last column).}
\tablefoot{
\tablefoottext{a}{Parameter frozen during the fit. }
\tablefoottext{b}{[keV]}
\tablefoottext{c}{[$10^{-2}$\,ph\,keV$^{-1}$\,cm$^{-2}$\,s$^{-1}$] }
\tablefoottext{d}{[$10^{-3}$\,ph\,cm$^{-2}$\,s$^{-1}$]}
\tablefoottext{e}{[atoms\,cm$^{-2}$]}
			}
 \label{tab:phav}
\end{table*}

The longest observations were obtained during the maximum of the outburst,
accordingly we focused on this data for the spectral analysis (observations
94032-04-02-03 to 94032-04-02-10).  
The spectra of 1A~1118$-$61 in different energy ranges were previously
described by power law and bremsstrahlung models \citep{maraschi,Coe:1994p3488}, but we found that
these models do not describe our data adequately. Our results
show that the broadband continuum of 1A~1118$-$61 can be well described by
the \texttt{FDCUT}, \texttt{NPEX}, and \texttt{CompTT} models
\citep{Coburn:2002p158,mihara,comptt} with qualitatively similar goodness
of the fit, modified by photoelectric absorption and iron emission line at
$\sim6.4$\,keV. 

All used continuum models in their best-fit residuals show a prominent 
absorption feature at $\sim55$\,keV. The inclusion of an absorption line
with a Gaussian optical depth profile leads to a significant reduction of
$\chi^2$ in all models (see Table~\ref{tab:phav}). We interpret this
absorption feature as a CRSF, which is detected in the spectrum of
1A~1118$-$61 for the first time. The observed energy of the feature is one
of the highest known and corresponds to a magnetic field B$\sim4.8\times10^{12}$\,G
\citep{harding} in the scattering region.
Preliminary phase-resolved analysis shows that the energy of the CRSF is
not likely to change significantly with the pulse phase (see
Fig.\ref{fig:enphase}).

The best-fit residuals of \texttt{FDCUT} and \texttt{NPEX} models show also a
prominent emission line-like feature at $\sim8$\,keV. Similar features at
different energies were reported for a number of sources (\citealt{Coburn:2002p158,rroca} and
references therein). \cite{Coburn:2002p158} suggested that the
employed phenomenological models for the continuum may be oversimplified
for the real sources and hence may be responsible for this effect. On the
other hand, \cite{Rothschild06} suggested that the feature may be
associated with the fluorescence copper line from the \textsl{Be/Cu}
collimator of the PCA instrument. To clarify the situation a more detailed analysis of
the \textsl{Suzaku} data is currently made and will be presented as a separate
paper (S. Suchy et al, in preparation). 

We found that the inclusion of an emission line with Gaussian profile may
help to account for this feature and does not significantly affect other
model parameters. The $\chi^2$ substantially improves for fits with
\texttt{FDCUT} and \texttt{NPEX} models, accordingly we included the line in all
fits with those models. The energy and the width of the line were fixed at
8.04\,keV and 0.01\,keV, corresponding to the copper $K_\alpha$ line, as proposed by
\cite{Rothschild06}. 

The fit may be slightly improved by the inclusion of an additional absorption
feature at $\sim23$\,keV for PCA data only. The investigation of the Crab
residuals (proposal P94802) shows that a similar feature is also present in
the PCA Crab spectrum (more than 99\% significance with maximum likelihood ratio (MLR) and F tests (\citealt{protassov}, p-value
$\sim0.006$). This is consistent with our 1A~1118$-$61 data, because
the line is not required by HEXTE data. On the other hand, a shallow
fundamental line at 22\,keV (with a first harmonic at 55 keV) could be more
difficult to detect due to this instrumental feature using PCA data.
A comparison of
our \textsl{RXTE} results with the \textsl{Suzaku} observations will help
to clarify the picture (S. Suchy et al, in preparation). 
\section{Conclusions} 
For the first time since 1992 a major outburst of the Be/X-ray binary
1A~1118$-$61 was observed in 2009~January with \textsl{RXTE}, following the
trigger of \textsl{Swift/BAT}. Strong pulsations with a period of
407.72\,s and a spin up of $-4.6\times10^{-7}$\,s\,s$^{-1}$ were
detected. A similar temporal behavior was observed with \textsl{CGRO}/BATSE
during the previous outburst. A
broadband spectrum of the source was obtained for a first time, and an
absorption feature at $\sim55$\,keV, interpreted as a CRSF was
detected. The inclusion of the feature significantly improves fit results
with all applied continuum models and its energy does not depend
significantly on the model used. 
More detailed spectral analysis is ongoing and the results will be
published elsewhere.

\begin{acknowledgements}
VD thanks DLR for financial support (grant 50OR0702).
\end{acknowledgements}

\bibliography{auto_clean}
\end{document}